\documentclass[twocolumn, aps, rmp, amsmath, amssymb, nofootinbib, superscriptaddress, longbibliography, floatfix, table-of-contents, eqsecnum]{revtex4-2}

\usepackage[pdftex]{graphicx}
\usepackage{mathrsfs}
\usepackage[colorlinks, breaklinks, urlcolor={blue}, linkcolor={red}, citecolor={blue}]{hyperref}
\usepackage{array}
\usepackage{amsmath}    
\usepackage{type1cm}
\usepackage{lettrine}
\usepackage[english]{babel}
\usepackage{lmodern}
\usepackage{microtype}
\usepackage{booktabs}
\usepackage[T1]{fontenc}

\hypersetup{
	pdfauthor = {Peter P. Rohde},
	pdftitle = {Quantum crypto-economics: quantum computing \& the crypto-economy},
	pdfsubject = {Quantum crypto-economics},
	pdfkeywords = {Quantum computing, Blockchain, Bitcoin, Cryptography}
}

\begin{document}

\title{Quantum crypto-economics: Blockchain prediction markets for the evolution of quantum technology}

\author{Peter P. Rohde}
\email{dr.rohde@gmail.com}
\homepage{www.peterrohde.org}
\homepage{www.keybase.io/peter\_rohde}
\affiliation{Centre for Quantum Software \& Information (QSI), University of Technology Sydney, NSW, Australia}

\author{Vijay Mohan}
\affiliation{RMIT Blockchain Innovation Hub, RMIT University, VIC, Australia}

\author{Sinclair Davidson}
\affiliation{RMIT Blockchain Innovation Hub, RMIT University, VIC, Australia}

\author{Chris Berg}
\affiliation{RMIT Blockchain Innovation Hub, RMIT University, VIC, Australia}

\author{Darcy Allen}
\affiliation{RMIT Blockchain Innovation Hub, RMIT University, VIC, Australia}

\author{Gavin Brennen}
\affiliation{Center for Engineered Quantum Systems, Dept. of Physics \& Astronomy, Macquarie University, 2109 NSW, Australia}

\author{Jason Potts}
\affiliation{RMIT Blockchain Innovation Hub, RMIT University, VIC, Australia}

\date{\today}

\frenchspacing

\begin{abstract}
Two of the most important technological advancements currently underway are the advent of quantum technologies, and the transitioning of global financial systems towards cryptographic assets, notably blockchain-based cryptocurrencies and smart contracts. There is, however, an important interplay between the two, given that, in due course, quantum technology will have the ability to directly compromise the cryptographic foundations of blockchain. We explore this complex interplay by building financial models for quantum failure in various scenarios, including pricing quantum risk premiums. We call this `quantum crypto-economics'.
\end{abstract}

\maketitle

\tableofcontents

\section{Introduction}


Quantum computing \cite{bib:NielsenChuang00} has become widely recognised as having the potential to compromise essential elements of present-day cryptographic techniques, especially public-key cryptography and digital signatures. This has profound implications for emerging technologies such as blockchain. There are many misconceptions, however, on how this could occur. We spell out the margins where quantum computing could impact blockchain. We also present a financial market predictor of the likelihood of a successful quantum computer attack on blockchain-based assets. We define a successful quantum attack on a blockchain as `quantum failure'.

%
%

As a note on terminology, we refer to the original blockchain described by \cite{bib:Satoshi} as the `Blockchain', and its associated cryptocurrency is `Bitcoin'. When we use the term `blockchain' we are referring to any generic blockchain. We discuss the Blockchain and blockchains in general for our purposes in the next section. For a discussion of blockchains in general see \cite{bib:Malekan} for an introductory coverage, \cite{bib:Werbach} for a more advanced overview, and \cite{bib:BergDavidson} for a discussion of the economic implications of the blockchain.

We argue that quantum failure can manifest in two important and distinct ways: first, as a purely monetary phenomenon that reduces the value of the native cryptocurrency, but keeps the integrity of the ledger intact, and second, as an accounting/technological phenomenon that undermines the integrity of the ledger itself, making the blockchain and its native cryptocurrency worthless. We treat (and model) these as two distinct problems associated with quantum failure.

Consider the monetary aspect of a quantum attack first. Quantum failure can allow an attacker to solve a computational problem faster than other miners (on average), thereby earning the majority of the block rewards over the length of time the attack persists. For a system such as the bitcoin Blockchain, this implies that mining can produce coins faster than the current 6.25 coins every 10 minutes (potentially until all 21 million feasible Bitcoins have been produced). We refer to this phenomenon as \emph{Grover-expansion}, because it increases the rate of monetary expansion (in this instance, of the native currency of a blockchain), a phenomenon well-known and well-understood in economics.

A couple of points are worth noting here. First, as defined here, Grover-expansion does not necessarily destroy the blockchain because if done on a small scale, it simply adds legitimate entries at a faster pace. Intuition would suggest that this monetary expansion will reduce the value of the native cryptocurrency, but need not necessarily reduce the value to zero. Second, many blockchain systems have a difficulty parameter built into the algorithms that oversee the rate at which these computations can be solved. In the case of Bitcoin, for example, the algorithm attempts to ensure that, on average, a block is added every 10 minutes. One could argue, therefore, that the difficulty parameter will adjust to negate the faster rate at which the quantum attacker solves the mining problem. The reality, however, is that the difficulty parameter is adjusted at discrete points in time, based on average computation rates in the past. For Bitcoin, the difficulty parameter is adjusted every 2016 blocks; at the current reward rate of 6.25 Bitcoin for every block, the attacker could earn a maximum of 12,600 Bitcoin before the parameter is adjusted. Moreover, once adjusted, the attacker would still be the fastest to solve the now more difficult computational problem.

If Grover attacks are done on a large scale, i.e by a large mining pool equiped with quantum computers, then even more dangerous attacks are possible like the 51\% attack. This occurs when the pool has over half the computational power of the network and allows dominating the blockchain. For example, such a dominant pool can perform a `double spend attack' by performing a spend transaction on one branch of the blockchain while growing a parallel branch where that spend record is missing. Given the computational dominance, this parallel chain will likely grow larger than the original and trusted nodes will adopt it, hence allowing for a second spend at no additional cost.

The second manner in which a quantum attacker can exploit the blockchain is by falsifying digital signatures and stealing existing tokens. This quickly erodes trust in the ledger and could elicit panic selling. 

We refer to this type of attack as a \emph{Shor-attack}; by rendering the blockchain entries unreliable, a Shor-attack would reduce the value of the native cryptocurrency, and indeed all assets denominated in that cryptocurrency, to zero. While this problem is distinct from Grover-expansion, it is quite possible that in some instances, depending on the nature of the blockchain, both attacks are launched simultaneously (or sequentially, with Grover-expansion preceding a Shor-attack), resulting in a Grover-Shor attack.

There are two major threat vectors enabled by a Shor-attack. The first is a fast steal. After a legitimate transaction has been added to the network but before it has been verified (usually within 10 minutes), the attacker can learn the private key of the sender from the publicly announced key. Then the attacker can broadcast a new transaction from the same sender's address to themselves. If the attacker offers a higher transaction fee, that transaction will take priority in the queue and will be verified first, meaning successful and unstoppable theft. The second is the recovery of lost Bitcoin. It is estimated up to 33\% of Bitcoin allocated so far on the network are associated with dormant public addresses from owners who presumably have lost their private keys and cannot access the coins \cite{bib:Stewart}. A Shor-attack would allow the attacker to learn the private key and take those coins. This clearly would increase available supply and devalue the currency especially if released quickly.

To understand the impact these attacks can have on financial markets, we construct a model that utilises simple bond pricing analytics, along with well-known parity conditions from exchange rate economics, to derive a financial indicator associated with the possibility of quantum failure. This model is useful for a number of reasons. First, it provides us with a method to infer the belief the market places on the probability of a quantum attack on a specific blockchain based on the risk premium of a bond denominated in the native cryptocurrency (or \emph{crypto-bond}) of the blockchain. Second, the simplicity of the model structure allows for a careful delineation of how different types of quantum attacks --- Grover-expansion and a Shor-attack --- impact the risk premium for a crypto-bond threatened with a quantum attack. Third, the use of familiar parity conditions yields intuitive insights on relationships between the crypto and fiat (say USD) economy. For example, it is readily shown that under certain simplifying assumptions, the percentage change in the exchange rate between the cryptocurrency and USD equals (approximately) the difference between the rate of Grover induced monetary expansion in the blockchain and the rate of inflation in the US economy; this is, of course, similar to purchasing power parity, and its derivation and intuition does indeed follow from that well-known exchange rate parity condition. Fourth, the model facilitates some simple comparative static exercises to gauge how parametric changes affect bond pricing analytics in this situation. Finally, the model sets the stage for a preliminary consideration of strategic issues relevant in this complex environment where quantum attacks that must be protected from \emph{ex ante} and defended against \emph{ex post}.

To motivate the final point regarding strategic interaction in this environment, we note that not all blockchains are equal --- some are more valuable than others. It is also unclear what the motivations of a quantum hacker may be. A hacker motivated by financial gain may attack a blockchain based on their perception of success and amount of financial gain to be had. An ideological hacker, however, may choose to attack a very different blockchain for very different reasons. Whatever their motivation, it is unlikely that a quantum hacker would command the resources to attack all blockchains at the same time. This implies the existence of a malicious quantum hacker would become known before all blockchains could or would be attacked. This in turn raises questions as to what the optimal response would be to this information? It is very clear that a research agenda should exist that explores all these issues and likely responses to quantum failure.

In Sec.~\ref{sec:blockchain}, we set out the features of blockchain that are important for our argument. We then explain the quantum computing challenge to blockchain in Sec.~\ref{sec:challenge}, and our financial model in Sec.~\ref{sec:model}. A conclusion and suggestions for further research follows in Sec.~\ref{sec:conclusion}.

\section{Blockchain \& cryptography}\label{sec:blockchain}

\subsection{Blockchain}

The Blockchain, which forms the basis for modern cryptocurrencies including Bitcoin, is a distributed data-structure which maintains a cryptographically irrevocable, chronologically-ordered ledger of transactions between agents.

The original blockchain (the Blockchain) was designed by Satoshi Nakamoto \cite{bib:Satoshi} to facilitate Bitcoin. This native internet currency has two important features of interest. First it avoids a double-spending problem and it solves the Byzantine generals' problem, which briefly put is the following: ``How do you assure that multiple distant parties agree on the same plan of action even in the presence of a small number of malicious traitors?''

 It performs both these functions by keeping an immutable record of the history of each Bitcoin. Economic agents --- known as miners --- are incentivised to maintain the Blockchain through the issuance of Bitcoin tokens as they add blocks containing transactions to the Blockchain. In the Blockchain protocol, miners spend computational effort to solve a puzzle (known as `proof of work') and the first one to do it is rewarded with a Bitcoin issuance as well as a transaction fee.
 
These Bitcoins are issued at a known, but diminishing, rate that was algorithmically established by Satoshi Nakamoto. In total there will be 21 million Bitcoin issued over time, with the issuance amount halved roughly every 4 years. It is useful to differentiate between the Bitcoin issuance rate and a monetary inflation rate. The issuance rate is the rate at which the number of tokens in a blockchain increase over time. An inflation rate is the rate at which a monetary unit loses value. High issuance of a monetary unit (popularly known as `printing money') is usually associated with high inflation rates. It is important, however, to maintain a distinction between these two concepts. A high rate of issuance of a blockchain token (Bitcoin being the original token) is not guaranteed to result in price inflation (devaluation of that token).

Importantly for our purposes, blockchain is based on several cryptographic primitives:
\begin{enumerate}
	\item Digital signatures are used to certify transactions under a consensus algorithm, whereby a sufficient number of independent parties must collectively agree to sign off on the legitimacy of a transaction for it to be transcribed to the ledger.
	\item Hashing is used to provide chronological links between connected transaction blocks, thus making the blockchain immutable, and Merkle tree data structures allow for fast and efficient verification of transactions by all parties \cite{bib:MerkleTree}.
	 \item The computational difficulty of inverse hashing is used in some proof-of-work protocols, such as Bitcoin, for mining new coins.
\end{enumerate}

These cryptographic primitives, however, have susceptibilities to the deployment of quantum computing, discussed in Sec.~\ref{sec:challenge}.





\subsection{Hash functions}

Hash functions, also an essential cryptographic primitive with widespread applications, that take an arbitrary input bit-string and reduce it to a short, fixed-length \emph{hash} that acts as a fingerprint of the original data in highly compact form,
\begin{align}
	h(x) \to y.
\end{align}
While the hash doesn't contain the original data, good (cryptographic) hash functions are designed to make it incredibly difficult to infer what the (or a) likely input was that it corresponds to, on the basis that they are \emph{one-way functions} (i.e computationally easy to evaluate in the forward direction, but extremely computationally difficult to invert) and exhibit highly quasi-random behaviour.

The problem of \emph{inverting hash functions} --- finding an input that maps to a given hash,
\begin{align}
	x = h^{-1}(y),
\end{align}
is, from a computational complexity perspective, conceptually similar to the problem of brute-forcing a symmetric-key crypto-system, and can be described mathematically in the same manner as a satisfiability problem. Note that, unlike the similarly-structured case of symmetric-key encryption, there is no key involved, and because the length of the output is fixed, there are necessarily \emph{collisions}, whereby multiple unique inputs map to the same output hash.

The Bitcoin Blockchain implements proof-of-work via \emph{inverse hashing}, whereby a legitimate coin is defined as a bit-string whose hash lies within a particular range. Specifically,
\begin{align}\label{eq:legit_coin}
	\text{SHA256}(\text{SHA256}(c)) = y
\end{align}
where SHA256 is a standard 256-bit hash function (256-bit Secure Hash Algorithm), and $c$ represents a legitimate coin when the output bit-string $y$ satisfies the constraint of having a fixed number of leading zeros. Since hash functions are one-way functions by definition, the mining process requires repeatedly hashing random bit-strings via brute-force until an input is found satisfying the output constraint.

In economic terms, this implies that monetary expansion is directly related to the \emph{hash rate} of computers allocated towards the mining process, modulated by a \emph{difficulty function}, which algorithmically ensures that mining becomes progressively harder by adjusting the number of necessary leading zeros in the output bit-string $y$ from Eq.~(\ref{eq:legit_coin}). This imposes an asymptotic cap on the token supply. Thus, the token issuance rate is effectively determined by the collective hash rate, and the algorithmically-imposed difficulty function.

\subsection{Public-key cryptography}

The most important cryptographic primitive we have today is public-key cryptography. Here the encryption-decryption process relies on two distinct keys, a \emph{public-key} which can \emph{only} be used to encrypt messages, and a \emph{private-key} which can \emph{only} be used to decrypt them. For this reason, such protocols are also referred to as \emph{asymmetric cryptography}. Both keys combined are jointly referred to as a \emph{key-pair}.

The encryption and decryption operations can be defined mathematically as,
\begin{align}
	f(m,k_\text{pub}) &\to c, \nonumber\\
	f^{-1}(c,k_\text{priv}) &\to m,
\end{align}
where $f$ and $f^{-1}$ are the encryption and decryption operations, $m$ is the plain-text message, $c$ is the cipher-text, and \mbox{$k=\{k_\text{pub},k_\text{priv}\}$} is the key-pair comprising the public and private keys.

The reason asymmetric cryptography is so useful, is that in most real-world scenarios in a globalised economy, we cannot go and meet people in advance of communicating with them to securely exchange secret keys. In fact, we may not even know them. By making everyone's public-keys openly available, we are able to securely send messages to them, without first having to perform any secret key-exchange with them a priori in a dark alleyway.

Importantly, one of the main criteria for such protocols is that the public and private keys cannot be computed from one another. Obviously, if one were able to look up a public-key on a key-server, and infer the associated private-key from it, the whole thing would be useless.

The original asymmetric cryptographic protocol was RSA, named after its inventors \cite{bib:RSA}. More recently, more efficient elliptic-curve cryptography (ECC) has become the norm \cite{bib:KoblitzECC, bib:MillerECC}.

The mathematical basis of RSA is that a private-key is defined by two prime numbers, and the corresponding public-key is given by their product. The computational security arises from the unproven, but widely held belief that integer factorisation is an extremely computationally complex problem --- believed to be \textbf{NP}-intermediate --- that classical computers cannot efficiently solve. This directly equates to their inability to infer private-keys from public-keys --- a so-called \emph{trapdoor function} --- and hence the cryptographic integrity of RSA. ECC is based on a different, but closely related mathematical problem, with the same essential cryptographic properties.

\subsection{Digital signatures} \label{sec:dig_sig}

While public-key cryptography can be utilised for securing messages, by reversing the roles of the public and private keys it can also be used to authenticate messages via the provision of \emph{digital signatures}. When you receive an email from someone, you want to be sure it was actually they who sent it, and that the email you received hasn't been modified or forged by someone else. Similarly, when logging into an online service like your email, users want to be sure they're interacting with whom they actually think it is.

Using the same public-key system as before, let everyone who wishes to be able to authenticate their messages create an additional key-pair. This time, they make the key that can only be used for decryption public, keeping the one that can only be used for encryption private.

Now when they send a message they want to digitally sign, they encrypt it (or just a hash of it), that they send along with the message. This encrypted hash acts as a digital signature, which others cannot falsify without knowing the privately-held encryption key. However, the publicly available decryption key can be used, upon receipt of a message along with its digital signature, to decrypt the signature and compare it with the message itself, thereby establishing its integrity.

\section{The quantum challenge to blockchain}\label{sec:challenge}

Large-scale quantum computers directly compromise both RSA and ECC, via a quantum algorithm known as Shor's integer factorisation algorithm \cite{bib:ShorFactor} and closely related discrete logarithm algorithm, allowing them to efficiently calculate private-keys directly from their associated public-keys. If and when quantum computing eventuates, this implies that the entire cryptographic backbone of our contemporary internet infrastructure would be compromised.

In addition to compromising encrypted messages, Shor-armed quantum computers could by the same logic also compromise digital signatures.


The analysis changes slightly, however, for symmetric cryptography (and similarly for inverse hashing). Here two parties share a secret-key in common, which is used for both encryption and decryption. Because the same key facilitates both roles, this is also referred to as \emph{symmetric} cryptography. The foremost current standard symmetric-key algorithm is AES256 (Advanced Encryption Standard, with 256-bit key-length) \cite{bib:Daemen}. This standard is a widely-used cryptographic primitive in today's internet software infrastructure.

In general, good symmetric cryptographic techniques are regarded as being very strong, in the sense that they are considered highly robust against conventional cryptanalytic techniques, such as differential cryptanalysis, which aren't known to provide significant shortcuts over brute-force attacks. It's therefore largely reasonable to associate their security with their vulnerability to brute-forcing, whereby we systematically try out all possible keys, attempting to decrypt an intercepted message using each one, and then running some simple tests (e.g statistical or language tests) to flag whether one such decryption attempt is likely to correspond to the unencrypted plaintext message (it's effectively certain that an incorrect key will not pass this test, providing a false-positive).

Using an encryption algorithm with key-length $n$, there are $2^n$ possible choices to work through, of which on average we'd have to try half until we find the right one. This exponential scaling grows extremely rapidly, and already $2^{256}$ combinations (as for AES256) far exceeds what any classical computer, present or future, could realistically iterate through systematically, from which the security of the algorithm arises.

This brute-force approach to cracking a code can be interpreted as what computer scientists refer to as a \emph{satisfiability problem} --- for some function that maps a plain-text input message and a key to an output cipher-text, \mbox{$f(m,k)\to c$}, which input value $m$ evaluates to a particular output $c$, assuming I don't know the key $k$? In this case, the input is a given choice of key, and the binary output answers the question `is this a legitimate decryption?'. In general, satisfiability problems are extremely computationally hard to solve. From the field of computational complexity theory \cite{bib:AroraBarak}, they are known to be \textbf{NP}-complete in general. Although it's unproven whether \textbf{NP}-complete problems can or cannot be efficiently solved on classical computers (proving this is one of the biggest open questions in the field of theoretical computer science), it is very strongly believed that they cannot be.

Formally, for a good cryptographic code, evaluating the function,
\begin{align}
	f(m,k) \to c,	
\end{align}
should be computationally easy, whereas evaluating its inverse,
\begin{align}
	f^{-1}(c,k) \to m,
\end{align}
should also be computationally easy if $k$ is known, but extremely computationally hard (or impossible) if it is not.


A quantum algorithm known as \emph{Grover's search algorithm} provides a relatively modest quantum advantage in solving this class of problems \cite{bib:Grover96}. The computational advantage it provides is to effectively quadratically reduce the input search-space over the equivalent classical brute-force approach. That is, whereas previously we had to search over $2^n$ possible input configurations, now we effectively have to search over only \mbox{$\sqrt{2^n} = 2^{n/2}$} of them. The right hand side of this equation provides the direct interpretation that it compromises security to the extent of effectively halving the respective key-length: $n$ becomes $n/2$. That is, the security of AES256 is effectively reduced to that of AES128. This is not an insurmountable problem, however, as a security response would be to simply switch to an AES512 standard.

Note that this enhanced scaling provided by Grover's algorithm is in the ideal context of error-free quantum computation. In reality, large-scale quantum computers necessarily require error-correction \cite{bib:Shor95}, which incurs significant overheads of its own (the scaling of which is highly architecture-dependent). For this reason, any practical future quantum advantage will be significantly less than the already modest advantage described above. Indeed, if the error-correction overhead is too great, there may be no advantage at all.

While symmetric cryptography is far more robust against quantum attacks than asymmetric cryptography, it cannot be used for digital signatures by virtue of its symmetry.

Although it appears that symmetric cryptography is robust to the advent of quantum computing, this is not proven. Security remains at the level of
\emph{computational security}, i.e the assumption that computers, classical or otherwise, are unable to provide sufficient computational resources to implement a systematic brute-force attack, or that future cryptanalytic techniques will not provide shortcuts around it.

A far stronger claim to security is via \emph{information-theoretic security}, whereby no such assumptions about adversarial computational capabilities are made. There is one (and only one) such symmetric-key algorithm which provides this information-theoretic level of security --- the \emph{one-time-pad} (OTP) algorithm, also known as the Vernam cipher. This algorithm, however, is absurdly impractical for use. For any given encrypted message, a key of exactly the same length must be employed, which cannot then be reused; hence the name.

Encryption is performed via a bit-wise XOR (or modulo-2 addition) operation between the plaintext message and key,
\begin{align}
	c = m \oplus k.
\end{align}
Decryption is performed by repeating the same procedure on the encrypted
cipher-text,
\begin{align}
	m = c \oplus k.
\end{align}
Thus the protocol is symmetric.

This `solution', however is trivial: If agents had the ability to share a key of the same length as the message itself, which couldn't subsequently be reused, they could use that opportunity to directly communicate the message.\footnote{The reason the key can only be safely employed once is because if two cipher-texts encrypted with the same key are XORed together we obtain the same as the XOR of the plain-texts, $c_{1} \oplus c_{2} = (m_{1} \oplus k) \oplus (m_{2} \oplus k) = m_{1} \oplus m_{2}$, upon which we can directly apply a conventional two-letter frequency attack to statistically predict $m_{1}$ and $m_{2}$. As soon as either of these are known we trivially extract the key via $k = m \otimes c$.} In short, one-time-pad encryption has very limited applicability.

Quantum cryptography provides a potential avenue to resolving the problems associated with the OTP protocol. By exploiting the randomness inherent in the measurement of quantum states, something imposed by the laws of quantum mechanics, it is possible to construct protocols for securely sharing long random bit-strings between remote parties. This is known as Quantum Key Distribution (QKD) \cite{bib:BB84, bib:Ekert91}. It is impossible for the shared random strings to be compromised via intercept-resend attacks. This provides only the ability to securely share random data not messages themselves. But by utilising this secure shared randomness as a source for OTP keys, it is possible, in principle, to resolve its impracticalities.

Of course, QKD need not only be used for OTP keys, but could be used for sharing secrets keys for any other symmetric-key algorithm, such as AES256. This creates a hybrid protocol, whereby the distribution of secret-keys is facilitated by QKD, but any unknown vulnerabilities in the underlying conventional crypto-system go unchanged.

In contemporary encrypted communications systems it is common to employ similar hybrid schemes combining symmetric and asymmetric elements, where a public-key system is used to share a shorter \emph{session-key}, that is subsequently employed in a symmetric cipher\footnote{The motivation for exchanging a session-key for use in a symmetric cipher, as opposed to directly communicating using an asymmetric cipher, is that the former are block-ciphers that map plain-text messages to  cipher-text messages of the same length, $|c| = |m|$, whereas the  latter induces significant space overheads in the cipher-text, $|c| \gg |m|$, thereby making it most space efficient to employ the latter once-off and the former thereafter.}. Typically this is performed using the Diffie-Hellman key-exchange protocol \cite{bib:DiffieHellman, bib:MerkleDH}. Such a hybrid QKD system effectively substitutes only the component associated with the exchange of session-keys.

While quantum algorithms do not entirely compromise hashing, they very much compromise digital signatures based on conventional public-key cryptographic techniques. This implies the ability for future quantum computers to transcribe fraudulent transactions to the Blockchain via falsifying the consensus, thereby undermining the integrity of the transaction ledger it maintains.

In the worst-case scenario, this could effectively invalidate the future value of transcribed contracts from the point in time at which such a Shor-enabled compromise of the ledger becomes viable.

This highlights the importance of `post-quantum cryptography' for future quantum-proof blockchain implementations. Already efforts are underway towards this goal. For example, the Quantum Resistant Ledger employs hash-based digital signatures for this purpose (note that hash-based signatures, despite being presumably robust against quantum attack vectors, have caveats of their own). And NIST has launched a project to identify and standardise post-quantum cryptographic protocols (\href{https://csrc.nist.gov/projects/post-quantum-cryptography}{csrc.nist.gov/projects/post-quantum-cryptography}).

In summary, Grover's algorithm can enhance hash-inversion as an instance of a satisfiability problem. This implies that quantum computation has the potential to distort the algorithmic token supply policy. We define the ability to artificially enhance monetary expansion as `Grover expansion'. This combined with Shor's algorithm ability to compromise digital signatures suggests that blockchain technology would be severely compromised by the advent of quantum computing. This ability of quantum computing to compromise the public key cryptography that underpins blockchain, either through Grover-expansion or a Shor-attack, or both, can be described as `quantum failure'. An analysis of quantum attack vectors on cryptocurrencies is presented in \cite{bib:TomamichelBlockchain}. One of the notable results there, is that when accounting for fault tolerance overheads, the timeline for a successful Shor-attack is much sooner, perhaps 10-15 years, than for a substantial Grover-attack.

\section{A financial market indicator of quantum failure}\label{sec:model}

In this section, we construct a simple model based on \cite{bib:BiermanHass75, bib:Yawitz77} to predict the impact of quantum failure on the assets in a crypto-economy.

\subsection{Crypto-bonds without quantum failure}

To keep the analysis intuitive, we consider a zero-coupon bond, $B$, denominated in some cryptocurrency, $X$. The bond has face value $A$, and time to maturity of one-year. We do not allow other maturities, although these could readily be incorporated to derive yield curves.

Risk in this model centres around the risk of quantum failure. While there is no equivalent of risk-free treasury bonds in the crypto-bond market, future bond face value payments can be guaranteed through smart contracts. Consequently, default risk can be eliminated through the requirement that 100\% of the borrowed funds be kept as collateral (possibly in assets denominated in a different currency, or even physical assets) in escrow via a smart contract. Thus, while expensive, smart contracts can eliminate the possibility of the idiosyncratic risk of default by the bond issuer. We assume that all agents are risk-neutral.

The first scenario we consider is one where either due to technological advancements in blockchain technology or lack of advancement in quantum ones, or both, a bond has no risk of quantum failure. Thus, in addition to zero idiosyncratic default risk, we impose zero systemic quantum risk. Given a yield to maturity (YTM) of $i$ and price $P$, we have,
\begin{align}\label{eq:1}
	P = \frac{A}{1 + i}.
\end{align}

\subsection{Quantum failure with a risk-free asset in same unit of account}\label{sec:4.2}

Now consider a hypothetical intermediate case where there are two bonds existing simultaneously denominated in the same cryptocurrency $X$:
\begin{itemize}
	\item $B$, which is risk-free as before.
	\item $\hat{B}$, which is susceptible to quantum attack.
\end{itemize}

This cannot happen in general, because either the entire $X$ network is susceptible to quantum failure in the form of a Shor-attack (in which case all $X$ denominated assets are at risk) or the entire network if free from risk of quantum failure. Since quantum risk is systemic, this case is unrealistic and hypothetical because it assumes that quantum risk is idiosyncratic to some bonds. However, it is a useful intermediate stage in the thought process.

Let \mbox{$0 < \rho < 1$} be the probability that quantum failure does not occur in the one-year period until maturity. We assume this belief is held by all market participants. The extremities of $\rho=0$ and $\rho=1$ are trivial. In the case of $\rho = 0$ quantum failure never occurs and the problem reduces trivially to the standard analysis. In the second instance, $\rho = 1$, quantum failure occurs with certainty, in which case no rational agent will hold risky crypto-bonds at any positive price. We assume that if quantum failure occurs for $\hat{B}$, it will pay zero to the holder of the bond.

For the risky asset we have, similar to before,
\begin{align}\label{eq:2}
	\hat{P} = \frac{A}{1 + \hat{i}}.
\end{align}
Arbitrage between the risky and risk-free bonds yields,
\begin{align}\label{eq:3}
	\hat{i} = \frac{1 + i}{\rho} - 1.
\end{align}

From this we can determine the bond risk premium that arises from the possibility of quantum failure,
\begin{align}\label{eq:4}
R &= \hat{i}-i \nonumber\\
&= \frac{(1+i)(1-\rho)}{\rho}.
\end{align}

This result essentially translates risk of quantum failure to its respective bond risk premium. So, if quantum risk were to be idiosyncratic, by looking at the risk-premium $R=\hat{i}-i$, we can infer the expected probability $1-\rho$ the crypto-market is placing on quantum failure occurring, providing a mechanism (for even non-market-participants) to infer quantum risk from a market-based indicator.

\subsection{A Shor-attack on blockchain $X$}

Now consider the existence of two separate systems, one of which is free from quantum risk and one that is exposed to quantum failure. It is not necessary that the former exists in the crypto world --- it could well be the fiat economy. The two requirements needed to conceptualise `risk-free' in this context are:
\begin{enumerate}
	\item There must be no idiosyncratic risk of default for the bond.
	\item The system must not be susceptible to quantum failure.
\end{enumerate}

Fiat can satisfy the first requirement through the issuance of government securities (treasury bills and bonds) and, even if public key cryptography fails, can satisfy the second requirement through the exclusive use of cash (notes and coins). A crypto-system can satisfy the first requirement through smart contracts, and the second only if the required technological advancements occur.

For the purposes of our analysis, there is one system (either fiat or crypto) using currency $X^{*}$, and one blockchain, $X$, that is susceptible to quantum risk in the form of a Shor-attack that will render its cryptocurrency worthless. The former has a risk-free bond, $B^{*}$; the latter's crypto-bond, which is susceptible to quantum failure, is $\hat{B}$ from Sec.~\ref{sec:4.2}, and will pay out zero if there is a Shor-attack. In the following we assume $X^{*}$ to be fiat, specifically USD.

Since these two systems rely on different units of account to price assets, there is a current spot exchange rate, $S$, that measures the price of 1 unit of $X^{*}$ in terms of $X$. Thus, a bond priced at $P$ units of $X$ is worth $P/S$ units of $X^{*}$, and so on.

Further, suppose the expected spot rate at the end of the one-year period when the bond matures is $S^{e}$. There are two possibilities:
\begin{enumerate}
	\item The exchange rate is fixed: $X$ is a stablecoin pegged to $X^{*}$ (USD), in which case $S^{e} = S$.
	\item The exchange rate is flexible: $S^{e} \neq S$, in general.
\end{enumerate}

Suppose exchange rates are flexible. To link the yield to maturity of the crypto-bond with risk of quantum failure to the yield to maturity of the risk-free treasury bond, we proceed via two steps. First, we compare two assets with identical risk attributes --- the (hypothetical) risk-free crypto-bond, $B$, denominated in $X$ (from Sec.~\ref{sec:4.2}) with the risk-free bond denominated in $X^{*}$. Second, we value the risky crypto-bond, $\hat{B}$ (denominated in $X$).

For the first step, if $i^{*}$ is the interest rate on $B^{*}$, assuming uncovered interest parity (UIP) holds,
\begin{align}\label{eq:5}
	1 + i = \frac{S^{e}(1 + i^{*})}{S}.
\end{align}

For an equilibrium where both $\hat{B}$ and $B^{*}$ are held, from Eqs.~(\ref{eq:1}) and (\ref{eq:5}), we have,
\begin{align}\label{eq:6}
\hat{i} &= \frac{S^{e}(1 + i^{*})}{\rho S} - 1, \\
{S}^{e} &= \rho S\left[ \frac{1 + \hat{i}}{1 + i^{*}} \right],
\end{align}
or in terms of percentages,
\begin{align}\label{eq:7}
{\dot{S}}^{e} &= \rho\left[ \frac{1 + \hat{i}}{1 + i^{*}} \right] - 1,
\end{align}
where,
\begin{align}
	{\dot{S}}^{e} \equiv \frac{S^{e} - S}{S}.
\end{align}

Thus, the equilibrium expected appreciation or depreciation of $X^{*}$ in terms of $X$ depends not only on the interest rate differential, but also on the probability of quantum failure. For any other spot rate expectation, given the yields to maturity, arbitrage possibilities will exist.

The risk premium on the risky crypto-bond becomes,
\begin{align}\label{eq:8}
	R &= \hat{i} - i^{*} \nonumber\\
	&= \frac{(1 + i^{*})(1 + {\dot{S}}^{e} - \rho)}{\rho}.
\end{align}

Compared to the previous result from Eq.~(\ref{eq:4}), there is an added element of foreign exchange risk, that is captured by the presence of ${\dot{S}}^{e}$.

In the case where $X$ is a stablecoin, ${\dot{S}}^{e} = 0$, and if we know $i^{*}$ (say the interest rate on USD 1-year treasury bond) and $\hat{i}$ (the yield to maturity on the risky bond), we can infer $\rho$, the market expectation for quantum failure. In the case of a flexible exchange rate, however, the market's beliefs about $\dot{S}^{e}$ would have to be estimated before such an inference about market perceptions of quantum failure can be made.

\subsection{The impact of expansion}

To investigate the impact of increases in the rate of coin issuance through Grover-expansion, we begin by noting that typically no cryptocurrency, including Bitcoin, acts as a unit of account for the purchase and sale of goods and services. Rather, as pointed out by \cite{bib:Bolt19}, vendors who accept payments in cryptocurrencies often simply convert a fiat price to the cryptocurrency price using an exchange rate $S$.

Assuming that the currency free from quantum risk, $X^{*}$ (USD), is the unit of account, this essentially implies that the Law of One Price holds. If so, for any good or service $\phi$, the price denominated in cryptocurrency $X$ is $p_{\phi} = p_{\phi}^{*}S$, where $p_{\phi}^{*}\ $is the price of $\phi$ denominated in $X^{*}$. Since this is true for every good, it follows that Purchasing Power Parity (PPP) holds for any aggregate measure of price levels (such as the Consumer Price Index) in any period $t$. Denoting the price levels in period $t$ as $C_{t}$ and $C_{t}^{*}$, PPP implies,
\begin{align}\label{eq:9}
	C_{t} = C_{t}^{*}S_{t}.
\end{align}

Consider now the transactions version of the quantity equation \cite{bib:Fisher}, used in \cite{bib:Bolt19} in the context of pricing cryptocurrencies, which we now suppose holds in cryptocurrency system $X$,
\begin{align}\label{eq:10}
	C_{t}T_{t} = M_{t}V_{t}.
\end{align}

Here,
\begin{itemize}
	\item $V_{t}$ denotes the velocity of cryptocurrency $X$ during period $t$.
	\item $T_{t}$ denotes the quantity of goods and services transacted on blockchain $X$ during period $t$.
	\item $C_{t}$ is the aggregate price level in blockchain $X$.
	\item $M_{t}$ is the quantity of tokens (money) issued in cryptocurrency $X$.
\end{itemize}

Eq.~(\ref{eq:10}) can now be expressed as,
\begin{align}
	\frac{C_{t}}{C_{t}^{*}}{(C_{t}^{*}T}_{t}) = M_{t}V_{t},
\end{align}
where $C_{t}^{*}T_{t}$ denotes the value of transactions on blockchain $X$ but denominated in currency $X^{*}$. Given that PPP holds, it follows from Eq.~(\ref{eq:9}) that,
\begin{align}\label{eq:11}
	S_{t} = \frac{M_{t}V_{t}}{C_{t}^{*}T_{t}}.
\end{align}
If PPP holds in every period, we have,
\begin{align}\label{eq:12}
	S_{t-1} = \frac{M_{t-1}V_{t-1}}{C_{t-1}^{*}T_{t-1}}.
\end{align}

\begin{figure}[!htbp]
	\includegraphics[width=\columnwidth]{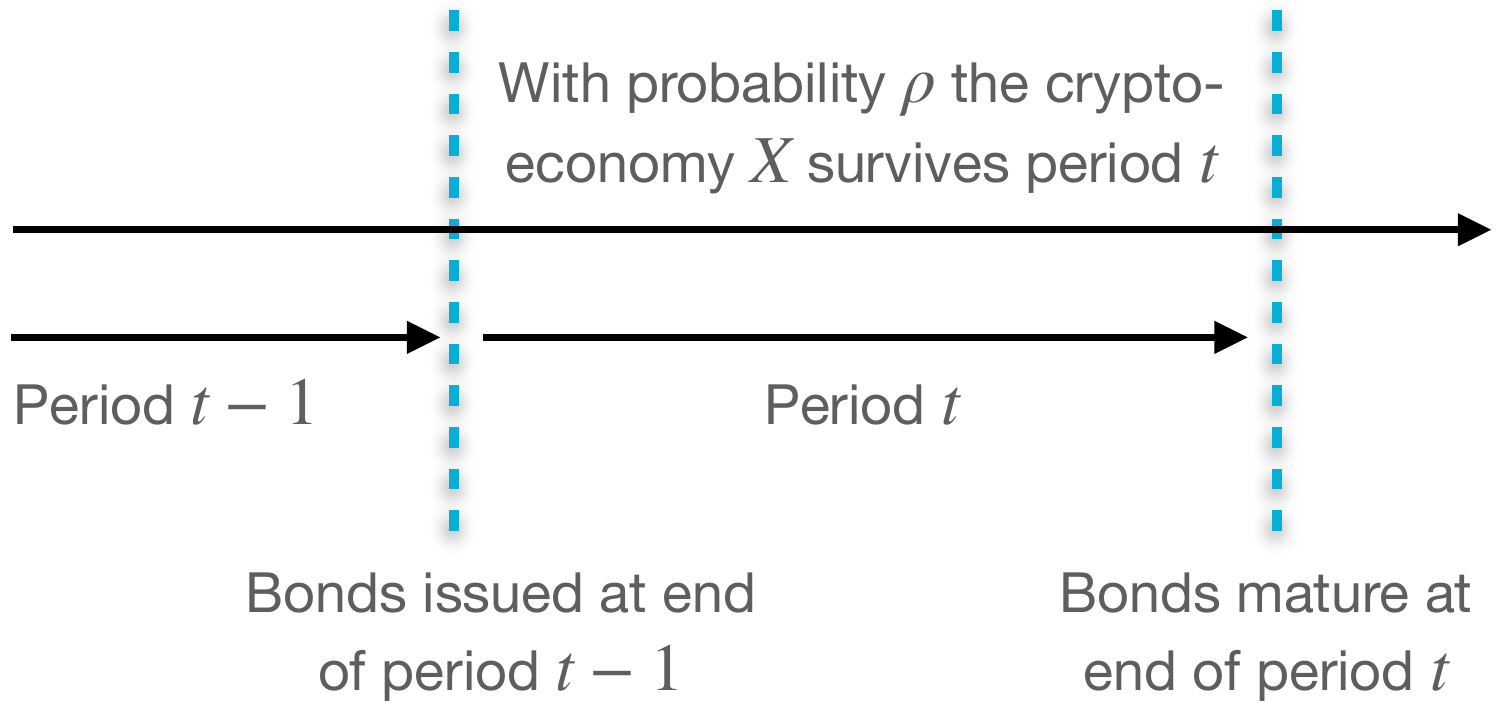}
	\caption{Timeline for the asset pricing model.} \label{fig:timeline}	
\end{figure}

To integrate this with the bond market analysis, it is worthwhile outlining the timeline more precisely, as shown in Fig.~\ref{fig:timeline}. Specifically, we assume that each period lasts for 1 year and we consider period $t-1$, as the past year. The bond is issued at the end of period $t-1$ (the current time) when the exchange rate is known to be $S_{t-1}$. The bond matures at the end of the following one-year period $t$.

As is evident from the timeline in Fig.~\ref{fig:timeline}, all the variables realised during period $t-1$ are pre-determined and known at the time of bond issuance (the end of period $t-1)$. Consequently, $S_{t-1}$, $M_{t-1}$, $V_{t-1}$, $C_{t-1}^{*}$ and $T_{t-1}$ are predetermined variables at the time of bond issuance and when investment decisions are made. At this time, moreover, investors must form expectations over the realisation of period $t$ variables. In the absence of a quantum attack, the issuance of money is algorithmically determined for cryptocurrencies, and $M_t$ is known to investors at the end of period $t-1$. Investors must form expectations over the remaining variables, which include: $S_{t}^{e}$, $C_{t}^{e*}$, $V_{t}^{e}$ and $T_{t}^{e}$. With this timeline, it follows that,
\begin{align}\label{eq:13}
S_{t}^{e} = \frac{M_{t}V_{t}^{e}}{C_{t}^{e*}T_{t}^{e}}.
\end{align}

Given the relatively short timeline we are focusing on (two periods, $t$ and $t-1$), let us assume that the velocity of the cryptocurrency is both stable \emph{and} common knowledge, such that $V_{t-1} = V_{t} = V$. Then, from Eqs.~(\ref{eq:12}) \& (\ref{eq:13}), we have, in the absence of quantum attack,
\begin{align}\label{eq:14}
\frac{S_{t}^{e}}{S_{t - 1}} = \frac{M_{t}}{M_{t - 1}}\frac{C_{t - 1}^{*}}{C_{t}^{e*}}\frac{T_{t - 1}}{T_{t}^{e}}.
\end{align}
Converting this to percentage changes gives,
\begin{align}\label{eq:15}
1 + {\dot{S}}^{e} = \frac{1 + \mu}{(1 + \pi^{*e})(1 + {\dot{T}}^{e})},
\end{align}
where:
\begin{itemize}
\item $\frac{M_{t}}{M_{t - 1}} = 1 + \mu$, where $\mu$ is the rate of change of money supply, or in other words, the rate at which new tokens are issued in $X$, which is an algorithmically determined constant.

\item $\frac{C_{t}^{e*}}{C_{t - 1}^{*}} = 1 + \pi^{*e}$, where $\pi^{*e}$ is the expected inflation of goods and services in the fiat system $X^{*}$.

\item $\frac{T_{t}^{e}}{T_{t - 1}} = 1 + {\dot{T}}^{e}$, where ${\dot{T}}^{e}$ is the rate at which transactions are expected to change over the given period.

\item $\frac{S_{t}^{e}}{S_{t - 1}} = 1 + {\dot{S}}^{e}$, where ${\dot{S}}^{e}$ is the expected appreciation/depreciation of $X^{*}$ in terms of $X$.
\end{itemize}

In order to focus on the role of token issuance on exchange rate expectations, let us assume that the volume of transactions using cryptocurrency is expected to be stable, i.e ${\dot{T}}^{e} = 0$. This assumption is readily dropped if the focus shifts from token issuance to gauging the impact of increasing or decreasing popularity of $X$ measured in terms of the volume of transactions, or if a more general approach is required. Under this assumption we obtain,
\begin{align}\label{eq:16}
1 + {\dot{S}}^{e} = \frac{1 + \mu}{1 + \pi^{*e}}.
\end{align}
The following approximation then holds (ignoring ${\dot{S}}^{e}\pi^{*e}$ for small percentage changes),
\begin{align}\label{eq:17}
	\dot{S}^{e} \cong \mu - \pi^{*e}.
\end{align}

Assuming that transaction volumes are stable (${\dot{T}}^{e} = 0$) and that the velocity of cryptocurrency $X$ is stable, the percentage change in the exchange rate is approximately equal to the rate of token issuance for cryptocurrency $X$ minus inflation of goods and services in the risk-free system $X^{*}$. So, for example, if BTC (blockchain $X$) token supply increases by 5\% every year and inflation of goods and services in the US (risk-free system $X^{*}$) is 2\%, we would expect the USD to appreciate by approximately 3\% in terms of BTC. As an aside, it is evident that the expectation of inflation in the US may itself be driven by the expected rate of change of the USD money supply by the Federal Reserve.

Now suppose we allow for the possibility of a quantum attack and Grover-expansion in period $t$. Letting $\mu^{G}$ represent the maximum possible monetary expansion that can be achieved by the quantum attacker by virtue of speeding up the mining process through enhanced computational (quantum) capabilities, Eq.~(\ref{eq:13}) transforms to,
\begin{align}\label{eq:18}
	S_{t}^{G} = \frac{M_{t}^{G}V_{t}^{e}}{C_{t}^{e*}T_{t}^{e}},
\end{align}
where,
\begin{align}
	\frac{M_{t}^{G}}{M_{t - 1}} = 1 + \mu^{G}.
\end{align}

Thus, allowing for the possibility of Grover-expansion the expected exchange rate is now,
\begin{align}\label{eq:19}
S_{t}^{e} &= \rho\frac{M_{t}V_{t}^{e}}{C_{t}^{e*}T_{t}^{e}} + \left( 1 - \rho \right)\frac{M_{t}^{G}V_{t}^{e}}{C_{t}^{e*}T_{t}^{e}} \nonumber\\
	&= \frac{V_{t}^{e}}{C_{t}^{e*}T_{t}^{e}}[\rho M_{t} + \left( 1 - \rho \right)M_{t}^{G}].
\end{align}
This yields the parallels to Eqs.~(\ref{eq:16}) \& (\ref{eq:17}),
\begin{align}\label{eq:20}
1 + {\dot{S}}^{e} &= \frac{1 + \rho\mu + \left( 1 - \rho \right)\mu^{G}}{1 + \pi^{*e}}, \nonumber\\
{\dot{S}}^{e} &\cong \rho\mu + \left( 1 - \rho \right)\mu^{G} - \pi^{*e}.
\end{align}

Eq.~(\ref{eq:20}) suggests that the expected exchange will incorporate the possibility of Grover-expansion in a quantum attack. Moreover, the higher the magnitude of Grover-expansion the market expects, the more the USD is expected to appreciate (and the cryptocurrency depreciate).

Apart from being interesting in its own right as a way to predict exchange rate changes based on algorithmically determined token increases and (fiat) inflation, this exchange rate forecast impacts the risk-premium of the risky bond if we combine combine Grover-expansion with a Shor-attack,
\begin{align}\label{eq:22}
R &= \hat{i} - i^{*} \nonumber\\
&= \frac{1 + i^{*}}{\rho}{\left[\frac{1 + \rho\mu + \left( 1 - \rho \right)\mu^{G}}{1 + \pi^{*e}} - \rho\right]}.
\end{align}

This implies that if $\hat{i}$ and $i^{*}$ are known along with the algorithmically determined $\mu$, the market expectation of the impact of Grover-expansion ($\mu^{G}$), and a publicly available forecast of expected inflation in system $X^{*}$, we can infer the market expectation of the probability of quantum failure, $1 - \rho$.

Finally, we can state some simple comparative static results. All else being equal, the risk premium on crypto-bond $\hat{B}$ increases when, from Eq.~(\ref{eq:22}):
\begin{enumerate}
	\item The rate of token issuance is higher (since \mbox{$\frac{\partial R}{\partial\mu} = \frac{1 + i^{*}}{1 + \pi^{*e}} > 0$}), unless deflation on goods and services in $X^{*}$ (say fiat) is more than 100\% ($\pi^{*e} < 0$ and $1 + \pi^{*e} < 0$).
	\item The rate of money supply increases due to Grover-expansion is higher (since $\frac{\partial R}{\partial\mu^{G}} = \frac{\left( 1 - \rho \right)(1 + i^{*})}{\rho(1 + \pi^{*e})} > 0$), unless deflation on goods and services in $X^{*}$ (say fiat) is more than 100\% ($\pi^{*e} < 0$ and $1 + \pi^{*e} < 0$).
	\item Inflation of goods and services in $X^{*}$ decreases (since $\frac{\partial R}{\partial\pi^{*e}} = - \frac{(1 + i^{*})(1 + \rho\mu + \left( 1 - \rho \right)\mu^{G})}{\rho{(1 + \pi^{*e})}^{2}} < 0$).
	\item And, of course, as the probability of $X$ surviving decreases (since $\frac{\partial R}{\partial\rho} = - \frac{(1 + i^{*})(1 + \mu^{G})}{\rho^{2}(1 + \pi^{*e})} < 0$).
\end{enumerate}

While we do not investigate further generalisations of the model, a number of extensions are possible, including: increasing the periods of analysis to generate a yield curve; increasing the maturity of the bond to allow for coupon payments; allowing for changes in the survival rate when there are more periods; investigating the role of demand through changes in transaction volumes, and so on. While these add greater sophistication to the model, they do not change the fundamental intuition behind the assessment of quantum risk when crypto-bonds are available.

\subsection{Quantum failure \& heterogenous blockchains}

Now suppose that there is one risk-free system (crypto or fiat), $X^{*}$, and $n$ blockchain systems $\{ X_{1},X_{2},\ldots,X_{n}\}$ that are at risk of quantum failure with associated survival rates of $\{\rho_{1},\rho_{2},\ldots,\rho_{n}\}$. Without loss of generality, suppose $\rho_{1} > \rho_{2} > \ldots > \rho_{n}$, such that the market believes $X_{1}$ has the greatest chance of surviving a quantum attack, and $X_n$ the lowest.

One question is why this belief structure exists. Given that quantum attacks are costly, the answer may depend on the attacker's objective. There are two separate issues here: first the security of blockchains vary; second, the value of blockchains vary. Moreover, `value' itself may depend on the attacker: a mercenary attacker may place the highest value on the blockchain with largest capitalisation; a political attacker may target a blockchain that causes the greatest harm to a certain group. Irrespective of their motive, the expected benefit (to the attacker) of an attack on $X_{i}$ is $\sigma_{i}v_{i}$, where $\sigma_{i}$ denotes the probability that an attack will succeed, and $v_{i}$ the (subjective) value to the attacker upon success.

For the market to form a clear assessment of $\{\rho_{i}\}$, agents may therefore first need to establish who the attacker is and their motives. That is, it is quite possible that the probabilities of quantum failure, $\{ 1 - \rho_{i}\}$, do not depend on technological security $\{\sigma_{i}\}$ alone.

Once such an assessment of $\{\rho_{i}\}$ is formed, however, the set of risk premia $\{ R_{i}\}$ reflect the market assessment of quantum failure in each system.

It is likely that in this situation, there being a single dominant blockchain --- for example, the state of affairs conceived by Bitcoin maximalists --- is not necessarily a good thing, because all the resources of the attacker could be devoted to compromising the security of the dominant system.

Indeed, this opens up a number of strategic possibilities open to attackers and market participants.

For example, suppose the attacker commences an attack on $X_{i}$. Assuming they do not have the resources to attack all $n$ systems, how should participants in other systems respond? Should they immediately abandon blockchains $X \setminus X_{i}$ and flee to the safety of $X^{*}$? But if that were the case, they could can destroy all $n$ systems by simply attacking the system with the greatest $\sigma_{i}$. The flight from all these blockchains will: (a) reduce their security to zero as miners depart; and, (b) drive the exchange rate value of all cryptocurrencies in terms of $X^{*}$, $1/S_{i}$, to crash (presumably to zero), making tokens worthless during the process of exchanging them to $X^{*}$.

A better market response might be to diversify portfolios across blockchains (possibly stochastically) such that mercenary attackers are less likely to find highly concentrated capitalisation worth attacking. This suggests some degree of randomisation (mixed strategies) both by agents and attackers. There are some interesting game-theoretic considerations here that may be worth exploring in future research.

\section{Conclusion \& future research directions}\label{sec:conclusion}

We have identified two vulnerabilities that blockchains have to the advent of quantum computing. Quantum hackers could falsify blocks being added to a blockchain and/or double spend tokens on any given blockchain depending on the features of the blockchain. This behaviour would manifest itself as monetary inflation --- a well-known and well-understood problem in the fiat economy. We are able to deploy standard economic analysis to develop a financial indicator that would reveal whether quantum failure had occurred or not. Our indicator relies on the existence of financially motivated individuals and pricing relationships that depend on the existence of efficient markets. As such we are confident that the indicator will be reasonably reliable in detecting quantum failure.

Our indicator, however, is only the beginning of an understanding of quantum failure. Scaleable quantum computing will not arrive spontaneously or immediately be deployed to looting blockchains or stealing wealth. Widespread adoption is likely to be gradual and, initially, can be expected to be expensive and possessed by few. In short, the initial applications are unlikely to be malicious. If the cost of dedicating quantum infrastructure to compromising a given cryptographic asset outweighs the realisable profit from doing so --- assuming financially motivated hackers --- a rational quantum-capable player would not be expected to make this investment. Thus, one would reasonably expect cheap cryptographic assets to retain their integrity via the lack of incentive to compromise them. On the other hand, cryptographic assets of tremendous value would logically stand higher in the list of priorities to quantum hackers.

These insights raise all manner of questions as to how best to respond to the generalised problem of quantum computers hacking blockchains. Which blockchains are likely to be hacked and by whom? How should observers react? Would a successful quantum attack result in a `flight to quality'? What does `flight to quality' mean? Is it a flight to fiat? Should individuals diversify their exposure to quantum risk across blockchains? If so, should the risk of quantum attack be thought of as being systematic risk as in Markowitz portfolio risk \cite{bib:Markowitz}?

Another avenue of study is the effect on crypto-economics when many adversarial agents are equiped with quantum computers. For example, in Grover's algorithm there is a non-negligible probability to solve a problem by measurement before the algorithm finishes. This gives rise to a mixed Nash equilibrium strategy for time to measure among many players trying to solve a problem like inverse hashing for mining first \cite{bib:Troy}. This would have implications on the fluctuations in transaction speeds, presenting new opportunities for market manipulation. 
	
Then there are geo-political risks. In addition to private agents developing quantum computing capability for their own purposes, nation states are also interested in quantum technology. How will this technology be regulated and controlled? The existence of rogue nation states developing capacity in this space is an immediate challenge to digital economic infrastructure over and above blockchain.

We propose that the analysis of the interplay between quantum computing and blockchain as economic infrastructure be labelled \emph{quantum crypto-economics}.

\section*{Acknowledgements}

Peter Rohde is funded by an ARC Future Fellowship (project FT160100397). Chris Berg, Sinclair Davidson, and Jason Potts are funded by an ARC Discovery (DP200101808). This research was fundedin part by the Australian Research Council Centre of Ex-cellence for Engineered Quantum Systems (Project number CE170100009)

\bibliography{bibliography}

\end{document}